\documentclass[12pt]{iopart}
\input epsf

\newcommand{\gl}[1]{Eq. (\ref{#1})} 
\newcommand{\gls}[2]{Eqs. (\ref{#1}) and (\ref{#2})} 
\newcommand{\beq}[1]{\begin{equation}\label{#1}}
\newcommand{\eeq}{\end{equation}}
\newcommand{\ttt}{{\tilde t}}
\newcommand{\omt}{{\tilde \omega}}
\newcommand{\mt}{{\tilde m}}
\newcommand{\qt}{{\tilde q}}
\newcommand{\kt}{{\tilde k}}

\begin{document}

\title{Colloidal gelation and non--ergodicity transitions
}
\author{J. Bergenholtz\\
}
\address{
Department of Physical Chemistry, G\"oteborg University, 412 96 G\"oteborg, 
Sweden\\
}
\author{M. Fuchs\footnote{To whom correspondence should 
be addressed} and Th. Voigtmann\\
}
\address{Physik-Department, Technische Universit\"at M\"unchen, 85747 Garching,
Germany\\\bigskip
}
\date{\today}

\begin{abstract}

Within the framework of  the mode coupling theory (MCT) of
structural relaxation, mechanisms and properties  of non--ergodicity
transitions in rather dilute suspensions of colloidal particles characterized
by strong short--ranged attractions  are studied.
Results building on the virial expansion  for particles with hard cores and 
interacting via an attractive square well potential  are presented, and their
relevance to colloidal gelation is discussed. 
\end{abstract}

\submitted{{\noindent \it }}

\section{Introduction}

Colloidal gelation  is a non--equlibirium transition
observed in dispersions where short ranged attractions exceeding a few $k_BT$
are present 
\cite{Carpinetti92,Bibette92,Poon97}.  It is accompanied by the formation of
(denser)  domains, which coarsen and finally freeze 
when the  gel line is traversed \cite{Poon93,Poon95}. 
The resulting gels are amorphous solids exhibiting finite elastic moduli
\cite{Grant93,Rueb98}. Often, the gel structure is fractal
 \cite{Carpinetti92,Poon95},
typically at low densities
\cite{Poulin99}, and often 
low angle scattering peaks are observed, which are reminiscent of spinodal 
decomposition; in this case gelation manifests itself by an arrest of the
time dependence of the peak position
\cite{Carpinetti92,Poon95,Verduin95,Verhaegh97,Poulin99}. 
This explicit dependence on the time since quenching the suspension shows that
colloidal gelation is a non--equilibrium phenomenon 
\cite{Poon97}.

Colloidal gelation has been observed in colloid polymer mixtures
\cite{Poon93,Poon95}, in solutions of sterically stabilized colloidal particles
if the solvent quality is decreased \cite{Grant93,Verduin95,Rueb98}, 
in emulsions \cite{Bibette92}, in solutions of charge stabilized particles
upon changes of the salt content \cite{Carpinetti92}, and in protein solutions
\cite{Muschol97}. In the last case, it prevents protein crystallization and
thus is an important 
obstacle for the collection of protein structure information 
\cite{tenWolde97,zukoski,Poon97c}.

Previous theoretical explanations of colloidal gelation have focussed on the
possibility of a dynamic percolation within a gas--liquid phase coexistence
region which is metastable with respect to gas--solid coexistence
\cite{Poon97}. The gel is assumed to form when the largest cluster spans the
sample and  the small angle scattering peak consequently arrests. 
Long ranged density fluctuations could arise from quenching below metastable
(\cite{Evans97}) spinodal lines as argued in Refs.
\cite{Bray94,Verduin95,Grant93,Verhaegh97}.
Fine--tuning of the attraction strength allows for the observation
of  all aspects of spinodal
decomposition in this systems without arrest and gelation
 \cite{Poon95,Poulin99}.
At low densities irreversible cluster--cluster aggregation provides a mechanism
explaining the observed small wave vector structures  \cite{Sciortino95}.
A review of experiments and theoretical approaches up to 1997 has been given by
Poon and Haw \cite{Poon97}.

Recently an alternative explanation for the arrest of the dynamics at 
colloidal gel transitions has been put forward
\cite{Bergenholtz99,Bergenholtz99b}. Non--ergodicity transitions triggered by
the local dynamics were suggested. Solutions of equations of the mode coupling
theory (MCT) for structural relaxation \cite{Goetze91b,gs} were studied which
exhibit a localization transition explaining a number of properties of
colloidal gelation. At low densities but for strong short--ranged  attractions,
particles are tightly bound to (asymptotically infinite) clusters and large
elastic moduli result \cite{Bergenholtz99,Bergenholtz99b}. 
In this limit, the MCT--equations can be simplified to one--parameter models 
which capture a number of the pertinent physical mechanisms although this
neglects other  aspects of the obtained colloidal gels  which are
connected to their self--similar structure on mesoscopic length scales.

In the following, the asymptotic models connected to colloidal gelation
shall be discussed further, where, especially, the existence of a divergent
cluster size  and its related dynamics is of interest.
The strength of the required attractions at gelation is estimated also.
Additionally,
two unphysical aspects of the previously obtained MCT results for  attractive
Yukawa potentials and for Baxter's adhesive hard sphere model (AHS)  
\cite{Fabbian99,Bergenholtz99} are reconsidered  using the controlled low 
density virial expansion \cite{hansen}. Adopting a very common model in colloid
science, hard spheres interacting with a square well potential are
studied \cite{russel}.

\section{Basic equations}

The most simple MCT equations for the normalized time--dependent intermediate
scattering functions at wave vector $q$,  $\Phi_q(t)$, shall be studied, as
appropriate for the description of the structural relaxation  of colloidal
suspensions (characterized by a Brownian short--time
diffusion coefficient $D^s_q$)
\cite{Bengtzelius84,Goetze91b,gs}:
\beq{e1}
\partial_t {\Phi}_q(t) + q^2 D^s_q \{ \Phi_q(t) + \int_0^t dt' \, m_q(t-t')
\; \partial_{t'} {\Phi}_q(t') \} = 0\; .
\eeq 
Here, the generalized longitudinal viscosity is given by a mode coupling
functional, $m_q(t) = {\cal F}_q([c],[\Phi(t)])$, which is uniquely specified
by the equilibrium static structure as given by the direct correlation function
$c_q$:
\beq{e2}
 {\cal F}_q([c],[f]) = \frac{\rho S_q}{2 q^4}\int 
\frac{d{\bf k}  d{\bf p}}{(2\pi )^3}\,
\delta({\bf k}\!+\!{\bf p}\!=\!{\bf q}) \,
({\bf q}\cdot{\bf p} c_{p}+{\bf q}\cdot{\bf k}c_k)^2
S_k S_p f_k f_p  \; .
\eeq
where $\rho$ is the particle number density and  $c_q$ determines the
static structure factor via, $S_q=1/(1-\rho c_q)$.

Non--ergodicity transitions of MCT dynamical equations are obtained as
bifurcation points of the non--linear equations for the
$q$--dependent non--ergodicity or Edwards--Anderson parameters $f_q$, which, in
the idealized MCT,   are defined as
$\Phi_q(t\to\infty)=f_q$:
\beq{e3}
 \frac{f_q}{1-f_q} = {\cal F}_q([c],[f]) \; .
\eeq
Whereas vanishing non--ergodicity parameters, $f_q=0$,
 indicate fluid or ergodic states,  finite ones, $f_q\ge
f^c_q>0$, spring into existence at larger interactions and signal the 
arrest of density fluctuations into non--equlibrium states, which are amorphous
solids like gels or glasses. The idealized MCT assumes a separation of time
scales describing the arrested structures as truly non--ergodic states,
although further slow transport processes may lead to a decay of the arrested
structures at much longer times; see the reviews \cite{Goetze91b,gs} for
discusions and extended theoretical approaches. 

Whereas glass transitions have been studied extensively and are present in the
studied colloidal suspensions at higher packing fractions also
\cite{Poon93}, non--ergodicity  transitions into gel
states exhibit novel phenomena and anomalies connected to the low packing
fractions and the resulting tenuous ramified gel structures. The experimentally
observed small--angle scattering peaks indicating domain formation 
after quenches into states far from equlibrium are one aspect, which however is
neglected here, because  short local length scales are considered where
local equilibrium may be assumed. Another aspect, the existence and growth of
tightly bonded clusters is signalled by the local dynamics; see below.

\section{Virial expansion input}

Specifying the static input, i.e. $c_q$,  the bifurcation lines and the
resulting long--time dynamics  can be
determined as functions of the thermodynamic control parameters, viz packing
fraction,  $\phi$,  and temperature, $T$.
For colloidal particles characterized by steric repulsion and short--ranged
attraction, the square well potential is widely used \cite{russel}:
\begin{equation}
u(r) = \left\{ \begin{array}{ll}
         \infty & 0 < r < \sigma  \\
 - U_0 
\;\;\; &  \sigma < r \le \sigma + \Delta  \\
0
\;\;\; &   \sigma + \Delta < r \; .\\
\end{array}
\label{sqw}
\right.
\end{equation}
The virial expansion provides a controlled approximation for dilute systems and
determines the direct correlation function for low packing
fractions as: $c(r) = f(r) +{\cal O}(\phi)$, 
where $f= e^{-u(r)/k_BT} - 1$ is the Mayr cluster function \cite{hansen}. The
gel transitions  can be studied in an asymptotic model 
of vanishing packing fraction and temperature, which describes a dilute system
of strongly interacting particles
\cite{Bergenholtz99,Bergenholtz99b}. Within the virial expansion this 
limit is obtained for:
\begin{equation}
\phi \to 0 \;\;\;\;\; \mbox{and}\;\;\;\;\; A\to \infty , \quad \mbox{so that}
\;\;\;\; \Gamma_v = \frac{12\Delta\phi A^2}{\pi^2\sigma}= \mbox{constant}\; ,
\label{asymp}
\end{equation}
where $A=e^{U_0/k_BT}-1$. Note that this results in $S_q=1$, because of
$\phi\to0$,  indicating
that structural correlations can be neglected in explaining the gel
non--ergodicity transitions. Thus, in this limit, glassy dynamics, due to
caging at higher packing fractions, is absent as well as dynamics connected
to critical phenomena, where long--ranged correlations are important; note that
an increase in the compressibility (or $S_{q=0}$) 
would require lower temperatures than
considered in \gl{asymp} as the virial estimate indicates:
 $S_{q\to0}\to\infty$ for $\phi A
\Delta/\sigma=1/24+{\cal O}(\phi,\frac\Delta\sigma)$.
Moreover, as expected on physical grounds, only attractive potentials can lead
to the limit \gl{asymp} and thus to gels, because $A$ would remain bounded for
pure repulsions; this is violated by the mean spherical approximation used in
\cite{Bergenholtz99,Bergenholtz99b}.

\section{Results for the gel structures}

The results for the gel form factors $f_q$ are of immediate interest
as they provide information on the spatial correlations of the arrested
solid--like structures.

\begin{figure}[h]
\centerline{{\epsfysize=8.cm 
  \epsffile{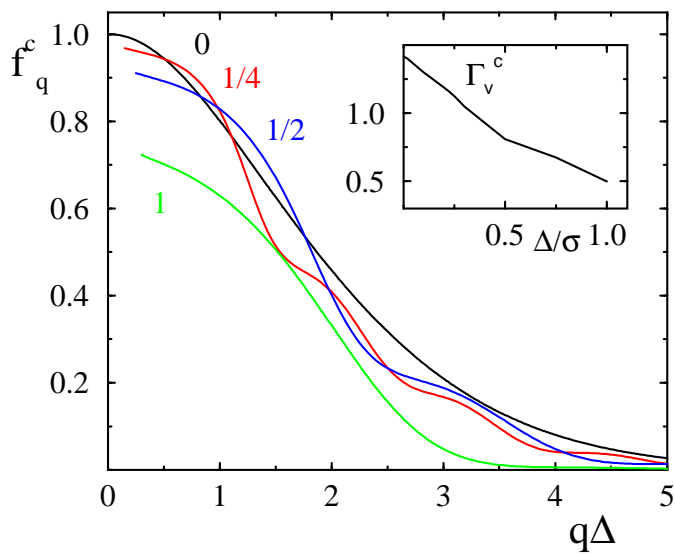}}}
\caption{Critical non--ergodicity parameter $f^c_q$  
of strongly interacting dilute fluids described  by Eq. (\protect\ref{asymp})
for the  well sizes $\Delta/\sigma=\frac 14$, $\frac 12$, $1$ and in the 
limit $\Delta\to0$;
most curves are not shown below $q\Delta=0.3$ because of 
numerical inaccuracies. 
The inset shows the variation of the critical coupling parameter $\Gamma_v^c$ 
with well size $\Delta$.
\label{fig1}} 
\end{figure}

In the specified limit of a strongly interacting dilute colloidal fluid, the 
mode coupling functional simplifies to a two parameter model as its density and
temperature dependence enters  via $\Gamma_v$ from \gl{asymp} only:
${\cal F}_q([c],[f])\to {\cal F}_q(\Gamma_v,\frac\Delta\sigma,[f])$.
Figure 1 shows non--ergodicity parameters at the transition points for three
different well sizes, $\Delta/\sigma= 1$, $\frac 12$, $\frac 14$, and in the
limit $\Delta\to0$. 
At the gel transition, bonds of a length given by the attraction range are
formed between  the particles; the well size $\Delta$ sets the length scale for
$f^c_q$. In the non--ergodic states, fast particle
rearrangements are possible on shorter distances,  causing the decay of
$\Phi_q(t=0)=1$ down to $f_q$, but  small wave vector
fluctuations are progressively suppressed upon decreasing $\Delta$.
Only for bond--lengths or well sizes $\Delta\ge0.1$, appreciable large distance
temporal fluctuations of the gel structure exist.

The additional limit $\Delta/\sigma \ll 1$ is
of interest, because the asymptotic model defined by \gl{asymp}  further
simplifies and the results from Fig. 1 indicate that this limit qualitatively 
captures the gel structures (i.e. $f^c_q$) for $\Delta$ below ca. 0.1.
Moreover, the elastic constants of the gel increase strongly in this limit of
short--ranged attractions, and thus observation of the non--ergodic gel states 
is more likely, as will be argued below. 

\begin{figure}[h]
\centerline{{\epsfysize=8.cm 
  \epsffile{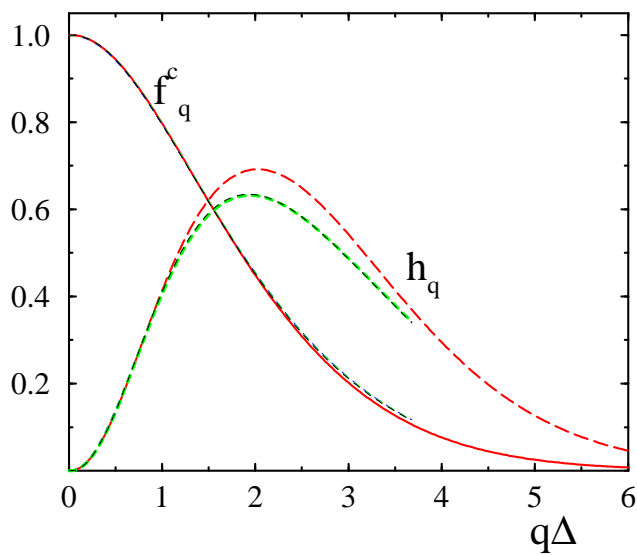}}}
\caption{Non--ergodicity parameter $f^c_q$ (full red line) and critical
amplitude $h_q$ (long dashed red line) at the critical point of the
 virial model for the 
square well potential. The thin short dashed
 black, green and blue lines  show the
corresponding results from the AHS model using the wave vector  cut--offs:
$k_{\rm max}\sigma=$ 40, 80 and  160; the effective width is found as
$\Delta^{\rm AHS}_{\rm eff.}=3.68/k_{\rm  max}$.
\label{fig2}} 
\end{figure}

A naive application of $\Delta\to0$ leads to the
AHS--virial result: 
$c_q \to 4 \pi A \Delta \sigma \frac{\sin{q\sigma}}{q}$.
Because of its slow algebraic decay  for $q\to\infty$, the wave
vector integration in \gl{e2} does not converge and the results depend on the
chosen wave vector cut--off $k_{\rm max}$. In the AHS--model this holds for all
temperatures and packing fractions as argued in \cite{Bergenholtz99}.
However, first entering the direct correlation function into \gl{e2}, and then
performing the limit $\Delta\to0$ leads to a different and 
almost everywhere convergent memory kernel:
\beq{model}
{\cal F}_\qt(\Gamma_v,[f]) 
= \frac{\Gamma_v}{\qt^2} \int d^3\kt\; 
 (\frac{{\bf \qt}\cdot{\bf \kt}}{\qt\kt^2})^2\; 
(1-\cos{\kt})\;  f_\kt\; f_{|{\bf \qt}-{\bf \kt}|}\;
\eeq
where $\qt=q\Delta$ denotes the rescaled wave vector. Tight localization of the
particles is predicted by the solutions to \gls{e3}{model} shown in Fig. 2, 
because the $q$--width of the Edwards--Anderson parameters $f_q$ is given by
the inverse of the narrow well width $\Delta$. Binding to in this limit
infinite clusters is described because  the wave vector dependent longitudinal
modulus diverges like $1/\qt^2$, even though it is connected to the total force
acting among all particles, $q^2 {\cal F}_q\propto\langle F_{\rm
tot.}^*(t\to\infty)F_{\rm tot.}(0)\rangle$  for $q\to0$, where the total force
among all particles vanishes according to Newton's actio--reactio principle,
$F_{\rm tot.}=0$.  In the specified limit, however, the particles experience
forces from infinitely removed particles belonging to the same cluster, and
thus $q^2{\cal F}_q$ stays finite as the true $F_{\rm tot}$ cannot be 
determined.
Collective and single particle density fluctuations consequently become
identical, $\Phi_q(t)=\Phi^s_q(t)$, leading e.g. to  the limit $f_\qt \to
1 - (\qt \tilde{r}_s)^2$ for $\qt \to 0$, where the particle root mean square
displacement, $r_s$, approaches (half) the well width,
$\tilde{r}^c_s=r^c_s/\Delta=0.48$,  at the transition point; see also
\cite{Bergenholtz99,Bergenholtz99b}. 

Figure 2 shows the solution of \gls{e3}{model} (evaluated with 400 grid points
spaced at $\delta\qt=0.025$)  at the critical point
$\Gamma_v^c=1.42\ldots$, which  corresponds
to rather modest attractions;  e.g. for $\Delta/\sigma=1/100$ and $\phi=0.1$
one finds $U_0/k_BT=3.56$. 
Solutions  from the AHS--virial approximation are also
included for three different cut--offs $k_{\max}\sigma=$ 40, 80 and 160. 
One notices that for small wave vectors the solutions can be scaled onto the
square well result; also the AHS critical coupling
parameter scales as one would  expect: $\Gamma_v^{c \, {\rm AHS}}\propto
1/k_{\rm max}$. However, as all wave vectors for
example enter the determination of the exponent parameter $\lambda$
\cite{Goetze85,Goetze91b,gs} appreciable differences result:
 $\lambda^{\rm AHS}_v=0.65$
compared to $\lambda^{\rm Sq.W}_v=0.79$.

\section{Results for the dynamics}

\begin{figure}[h]
\centerline{{\epsfysize=8.cm 
  \epsffile{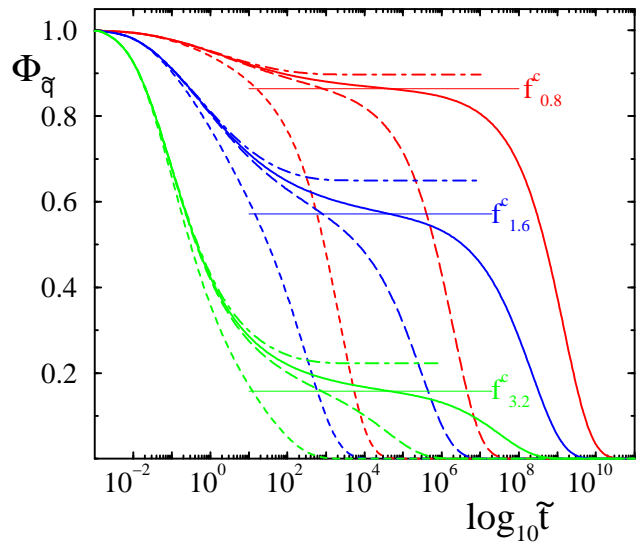}}}
\caption{Intermediate scattering functions $\Phi_\qt(\ttt)$ determined from 
(\protect\ref{e7}).
For the  interaction parameters $\Gamma_v/\Gamma_v^c= 0.9$, $0.99$, $0.999$,
and $1.01$, corresponding to the short--, long--dashed, solid and  
dashed--dotted curves, solutions for the  wave vectors $\qt=0.8$ (red), 
$1.6$ (blue) and $3.2$ (green)
are shown. The thin horizontal lines indicate the critical 
non--ergodicity parameters $f^c_\qt$ which give the long--time limits,
 $f^c_\qt=\Phi_\qt(\ttt\to\infty)$ at $\Gamma_v^c=1.3647$ for the chosen 
discretization, $\delta\qt=0.08$ and 100 grid points. 
\label{fig3} }
\end{figure}

The bifurcations to non--ergodic solid--like states in \gls{e3}{model}
also lead to anomalies in the long--time dynamics which provide the most
detailed information about the gel--formation and the connected transport 
mechanisms. We focus on the limit of narrow attractions, 
$\Delta\ll\sigma$, as clustering is specific to the considered low--density 
transitions, and it is thus of interest to study the consequences of 
the small--$q$ divergence of the longitudinal modulus. 

In the limit given by \gl{asymp} and for $\Delta\ll \sigma$, introduction of 
a rescaled time, $\ttt =tD_0/\Delta^2$ with $D_0=D^s_{q\to\infty}$,
 eliminates the transient parameters
and leads to simplified equations of motion for the coherent (and identically 
for the incoherent) density fluctuations:
\beq{e7}
\partial_\ttt {\Phi}_\qt(\ttt) + \qt^2 \Phi_\qt(\tt) + \int_0^\ttt d\ttt' \; 
\mt_{\qt }(\ttt-\ttt')
\; \partial_{\ttt'} {\Phi}_\qt(\ttt')  = 0\; ,
\eeq 
where the friction function is given by: 
$\mt_\qt(\ttt)=\qt^2 {\cal F}_\qt(\Gamma_v,[f])$, and  the initial variation 
is  $\Phi_\qt(t)=1-\qt^2\ttt+\ldots$.

Figure 3 shows intermediate scattering functions for a range of interaction 
parameters and wave vectors. The two--step relaxation can be analyzed with the
techniques developed for MCT--glass transitions \cite{Goetze91b,gs}. It
exhibits the factorization property during an intermediate time window:
$\Phi_\qt(\ttt) = f^c_\qt + h_\qt G^\lambda(\ttt)$ 
for $|\Phi_\qt-f^c_\qt|\ll1$, which
enlarges upon approaching $\Gamma_v^c$. The $\beta$--correlator 
$G^\lambda$, captures
the sensitive dependences on time and interaction strength, and the amplitudes
$f^c_\qt$ and $h_\qt$ (see Fig. 2) describe the gel structure and the spatial 
correlations of the mechanism dominating the bond--formation. 
Power--law decays around the plateau value $f^c$ and the  divergence of a
first scaling time are contained in $G^\lambda(\ttt)$.

The near arrest of the $\Phi_\qt(\ttt)$ around $f^c_\qt$ leads to an 
intermediate elastic behavior of the gel, which in the studied limit follows 
from the asymptotic result for the shear modulus $G$:
\beq{shear}
G(\ttt) = (k_BT\rho) \,  \frac{\pi\Gamma_v\sigma^2}{20\Delta^2}
  \int_0^\infty d\kt\; (1-\cos{\kt})\, \Phi_\kt^2(\ttt)\; .
\eeq
Figure 4 shows the corresponding storage and loss moduli which,
in the fluid, exhibit a near 
elastic plateau with power--law corrections predicted by the 
$\beta$--correlator 
$G^\lambda$. Using the initial variation of $\Phi_\qt(\ttt)$, a high--frequency
divergence, $G'(\tilde{\omega})\propto G''(\tilde{\omega}) \propto 
\tilde{\omega}^{1/2}$ for $\tilde{\omega}\to\infty$
results from the integration in \gl{shear}  at high $k$. 
This result, which is familiar  for colloidal hard spheres 
in the absence of hydrodynamic interactions \cite{russel}, 
cannot be seen in Fig. 4 because of the chosen small  wave vector 
cut--off.

\begin{figure}[h]
\centerline{{\epsfysize=8.cm 
  \epsffile{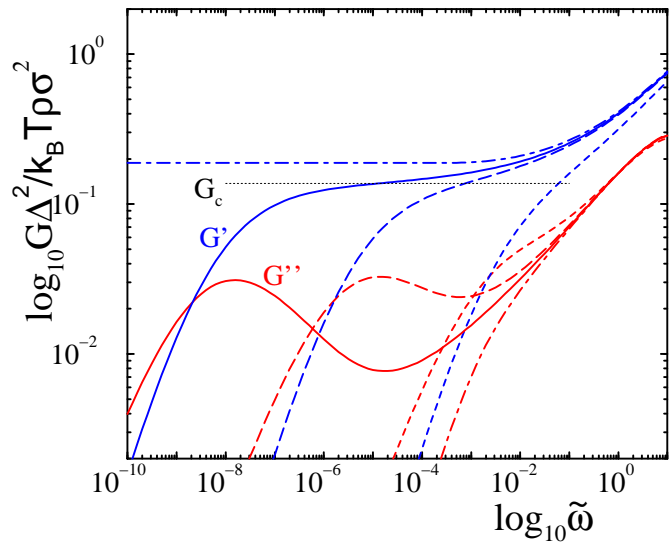}}}
\caption{Storage, $G'(\omt)$ (blue), and loss, $G''(\omt)$ (red),
 moduli according to Eq. (\protect\ref{shear}) for the interaction
 parameters and line styles  of Fig. 3. The horizontal line indicates the 
critical elastic modulus, $G_c$, of the gel at the transition, $\Gamma_v=
\Gamma_v^c$.  
\label{fig4} }
\end{figure}

Importantly, \gl{shear} predicts that large, ${\cal O}((\sigma/\Delta)^2)$,
elastic
moduli result upon formation of the gels for $\Gamma\ge\Gamma_v^c$,
 if the particles
interact via short--ranged attractions \cite{fehler}. 
The elastic modulus at the transition, $G_c$, which follows from \gl{shear} and
the critical gel structure factors, $f^c_\qt$, is included in Fig. 4.

The final decay of the intermediate scattering functions or of the shear 
modulus  $G$ is described asymptotically by the second MCT scaling law, 
which entails  the existence of a second set of divergent time scales and of
$\Gamma_v$--independent and  non--exponential relaxation functions
\cite{Goetze91b,gs}. The small--$q$ divergence of the longitudinal modulus
leads to a diffusive behavior of the final relaxation times, 
$\tilde{\tau}_\qt\propto \tilde{\tau}^\infty_\qt 
(\Gamma_v^c-\Gamma_v)^{-\gamma}$, with the $\Gamma_v$--independent amplitude
$\tilde{\tau}^\infty_\qt \propto 1/\qt^2$. Here the exponent 
$\gamma$ is determined by $\lambda$, and follows as $\gamma=2.85$.
The diffusive  behavior is apparent in Fig. 3 and stresses that 
the final melting of the precursor gel structure in the fluid phase requires
transport over large distances which scale like $\sigma/\Delta^2$ as follows 
from \gl{model}. Again this indicates the presence of 
infinite clusters which arrest and cannot melt anymore at and above the
gel transition at $\Gamma_v^c$.

\section{Conclusions}

We have provided further support for the suggestion that colloidal gelation is
caused by  non--ergodicity transitions triggered on local distances
\cite{Bergenholtz99,Bergenholtz99b}.

It was shown that non--ergodic structures characterized by strong elastic
moduli are obtained in dilute solutions of colloidal particles interacting
with short--ranged attractions upon lowering the temperature. 
Bonds with a length given by the attraction range are formed between the
particles if  the strength of the attraction increases 
to a few $k_BT$. These values are in qualitative agreement with
phenomenological considerations for reversible gelation
\cite{Verhaegh97,Poon97}. 
We suggest, that this formation of long--lived bonds is the rate limiting step
for (transient) gelation in colloidal suspensions.

The formation of  (percolated) clusters is indicated by the
long--ranged force correlations of the arrested structures which cause the
small wave vector divergence of the mode coupling functional ${\cal F}_q$.
The  description of the mesoscopic gel structure and of the related
domain coarsening kinetics, however, has not yet been incorporated into the
present approach. 

The long--ranged force correlations at the gel non--ergodicity transitions
cause that collective and single particle dynamics become
identical for long times, and  that one mode coupling functional determines all
time dependent small wave vector quantities, like mean square displacement
or  moduli; such connections have
already been exploited to measure the elastic response via light scattering
techniques \cite{Mason95}. 
Moreover, a diffusive final relaxation of the colloid density
fluctuations results close to the gel transitions, $\tau_q\propto 1/q^2$, which
suggests that a generalized Gaussian description, $\Phi_q(t)=\exp{\{-(q^2/6)
\delta r^2(t)\}}$,  as suggested by Segr{\`e} and Pusey \cite{segre96}, can
provide  a reasonable description of the long time dynamics for all wave
vectors; note that for colloidal suspensions at the glass transition this
ansatz failed at small wave vectors because of the non--diffusive structural
relaxation \cite{segre96,fuchsmayr}.

Using the exact low density virial expansion, 
two at first sight
unphysical aspects of previous MCT calculations were identified as artefacts
of the approximate static structural  input.
First, the cut--off dependence found within the AHS model
\cite{Fabbian99,Bergenholtz99}  was analyzed for low densities.
The interesting conclusion  is that MCT predicts the AHS model to be in the
non--ergodic state for any non--zero particle concentration. 
Second, it was clarified that gelation requires attractive interactions.

Dynamic light scattering measurements as preformed by van Megen and coworkers
at the colloidal glass transition \cite{Megen93,Megen94b,Megen98}
or further viscoelastic measurements \cite{Rueb98,Mason95} can provide crucial
tests of our approach. First measurements of the non--ergodicity parameters
support it \cite{Poon99}. 
Dynamic light scattering measurements of Krall and
Weitz \cite{Weitz98} also show intriguing connections and require further
 study.

\section{Acknowledgments}
\label{acknowledgements}

Valuable discussions with 
Dr. P. N.  Segr{\`e}, Professor D. A.  Weitz, Professor M. E. Cates and
Professor W. G{\"o}tze are gratefully acknowledged. 
M. F.  was supported by the Deutsche
Forschungsgemeinschaft under Grant No. Fu 309/2.

\section{References}


\end{document}